\documentclass[12pt,letterpaper]{JHEP3}

\usepackage{amsmath,amssymb}

\font\tenscr=rsfs10 scaled1100
\font\sevenscr=rsfs7 % scaled \magstep1 
\font\fivescr=rsfs5 % scaled \magstep1 
\skewchar\tenscr='177 
\skewchar\sevenscr='177
\skewchar\fivescr='177 
\newfam\scrfam 
\textfont\scrfam=\tenscr
\scriptfont\scrfam=\sevenscr 
\scriptscriptfont\scrfam=\fivescr
\def\linebreak{\hfill\break}

\def\bra<#1|{\langle #1\rvert}
\def\ket|#1>{\lvert#1 \rangle}
\def\braket<#1|#2>{\langle #1|#2 \rangle}
\def\const{\text{const}}

\def\inpare#1{\left(#1\right)}
\def\bigpare(#1){\left(#1\right)}

\def\insbra#1{\left[ #1 \right]}
\def\bigbra[#1]{\left[ #1 \right]}
\def\prefix#1#2{\,{}^{#1}\!{#2}}

\def\therefore{\mbox{\setbox0=\hbox{X}\hbox{$\ldotp$}\raise0.7\ht0\hbox{$\ldotp$}\hbox{$\ldotp$}} \quad }
\def\because{\mbox{\setbox0=\hbox{X}\raise0.7\ht0\hbox{$\ldotp$}\hbox{$\ldotp$}\raise0.7\ht0\hbox{$\ldotp$}}\kern0pt }

\def\SO{{\rm SO}}

\def\maps{\rightarrow}

\def\Frac(#1/#2){\left(\frac{#1}{#2}\right)}

\def\AdS{{\rm AdS}}

\def\Eq#1{\begin{equation} #1 \end{equation}}

\def\Eqr#1{\begin{eqnarray} #1 \end{eqnarray}}

\def\Eqrsub#1{\begin{subequations}\Eqr{#1}\end{subequations}}
\def\Eqrsubl#1#2{\begin{subequations}\label{#1}\Eqr{#2}\end{subequations}}
\def\Bitm{\begin{itemize}}
\def\Eitm{\end{itemize}}
\def\Blist#1#2{\begin{list}{#1}{\parsep=0pt \itemsep=0pt%   
  \listparindent=0pt #2}}
\def\Elist{\end{list}}
\long\def\ignore#1#2{\def\ignoreflag{#1}\long\def\tmptext{#2}
  \ifnum\ignoreflag>1 #2 \fi}

\newcommand{\nn}{\nonumber}
\newcommand{\pd}{\partial}

\def\slash#1{\ooalign{\hfil/\hfil\crcr$#1$}}

\def\Xsp{{\rm X}_4}
\def\Ysp{{\rm Y}_6}
\def\Tsp{{\rm T}^{11}}
\def\Zsp{{\rm Z}_5}

\numberwithin{equation}{section}

\title{Moduli Instability in Warped Compactifications 
of the Type IIB Supergravity}

\author{
Hideo Kodama and Kunihito Uzawa\\ 
Yukawa Institute for Theoretical Physics\\
Kyoto University, Kyoto 606-8502, Japan.\\
E-mail: \email{kodama@yukawa.kyoto-u.ac.jp, 
uzawa@yukawa.kyoto-u.ac.jp }}
\abstract{%
We show that the conifold and deformed-conifold warped 
compactifications of the ten-dimensional type IIB supergravity, 
including the Klebanov-Strassler solution, are dynamically unstable 
in the moduli sector representing the scale of a Calabi-Yau space, 
although it can be practically stable for a quite long time in a 
region with a large warp factor. This instability is associated with 
complete supersymmetry breaking except for a special case and 
produces significant time-dependence in the structure of the 
four-dimensional base spacetime as well as of the internal space.}
\preprint{YITP-05-20 \hfill\\{\tt hep-th/0504193}}
\begin{document}

%T1>Introduction
%======================================%
%<<<<<<<<<<<< INTRODUCTION >>>>>>>>>>>>%
%======================================%

\section{Introduction}
With the practical confirmation of the inflationary universe 
scenario of the early universe and the discovery of the accelerating 
expansion of the present universe
\cite{Riess:1998cb,
Perlmutter:1998np,Bennett:2003bz,Spergel:2003cb,Peiris:2003ff}, 
it is now the most challenging 
problem to construct a consistent cosmological model that explains 
these observational facts, on the basis of supergravity and string 
theory, which are the only viable unified fundamental theories at 
present. The main obstacle to this problem is the fact that these 
theories require the spacetime to be higher dimensional; in order to 
obtain a four-dimensional universe at low energies, we have to find 
a natural way to conceal extra dimensions, which is usually called a 
compactification. This compactification gives rise to various new 
problems. One of the most serious problems is the moduli 
stabilisation \cite{Kachru:2002he}. 
Another is the no-go theorem against accelerating 
expansion of the universe in simple Kaluza-Klein-type or stationary 
warped compactification with a smooth compact internal space
\cite{Gibbons:2001wy}.

Recently, a new progress in resolving these problems has been made 
by Kachru et al \cite{Kachru:2003aw,Kachru:2003sx}. 
Utilising a conifold-type flux compactification of 
the IIB supergravity that realises the stabilisation of all complex 
moduli \cite{Kachru:2003aw,Klebanov:2000hb}, they proposed a model in 
which all moduli are potentially stabilised and an accelerating 
expansion of the universe is realised for a sufficiently long time. 
There is, however, one subtle weak point in their model. It is the 
stability of the moduli degree of freedom corresponding to the scale 
of the Calabi-Yau internal space \cite{Giddings:2001yu}. 
They argued that this degree of 
freedom would be stabilised by quantum nonperturbative 
effects \cite{Witten:1996bn,Kachru:2000ih,Tripathy:2002qw}. However, their 
argument is based on a four-dimensional effective theory that does 
not take into account the warping and assumes the supersymmetric 
background.  

The main purpose of the present paper is to analyse the stability of 
this moduli degree of freedom in the classical framework of the 
full ten-dimensional IIB supergravity, taking account of the 
warped structure.  We show that many of well-known supersymmetric 
compactifications of the type IIB supergravity by a conifold 
\cite{Klebanov:1998hh,Klebanov:1999rd,Klebanov:2000nc}, 
a resolved conifold \cite{PandoZayas:2000sq} 
or a deformed conifold \cite{Klebanov:2000hb,Minasian:1999tt,Ohta:1999we} 
with 5-form flux and 3-form 
flux are limits of dynamical solutions with four extra parameters. 
This instability occurs in the warp factor in the form $h=h_1(y)+ 
a_\mu x^\mu + b$, where $x^\mu$ and $y^p$ are the coordinates of the 
four-dimensional base spacetime and the Calabi-Yau internal space, 
respectively, and for $a_\mu=b=0$, the solution reduces to the 
original supersymmetric one. 

In the special case in which there is no 3-form flux, the 
corresponding solution is identical to the one recently found by 
Gibbons, L\"u and Pope \cite{Gibbons:2005rt} as a higher-dimensional 
analogue of the four-dimensional Kastor-Traschen 
solution \cite{Kastor:1992nn}. In the case of a deformed conifold 
with 5-form and 3-form fluxes, the corresponding solution provides a 
time-dependent extension of the Klebanov-Strassler 
solution \cite{Klebanov:2000hb}.  We show that 
supersymmetry is fully broken, except in the case the solution has a 
null Killing, for the Gibbons-L\"u-Pope solution. 

The present paper is organised as follows. First, in Section 
\ref{sec:KS-solution}, after making clear ansatz to be imposed on 
various fields and the spacetime metric of the IIB supergravity to 
find solutions, we show that all field equations including the 
Einstein equations can be reduced  under the ansatz to a simple set 
of equations on the Calabi-Yau manifold for the 2-form potential 
$B_2$ and the warp factor $h$. We also show that $h$ is restricted 
to the form given above in general. Then, in Section 
\ref{sec:conifold}, we apply this formulation to various 
compactification models and derive explicit expressions for the 
solutions. In Section \ref{sec:susy} we study the supersymmetry 
breaking for the Gibbons-L\"u-Pope solution and discuss its 
implication. Finally, Section \ref{sec:discussion} is devoted to  
summary and discussion. 

%T1>TD solutions in IIB
%======================================%
%<<<<<<<<<<<<< SECTION 2 >>>>>>>>>>>>>>%
%======================================%

%\section{Time-dependent solution in the type IIB supergravity}
\section{Formulation}
\label{sec:KS-solution}

%T2>Ansatz
\subsection{Ansatz}

In the present paper, we look for solutions whose spacetime metric 
has the form
%============< EQUATION >==============%
%
\begin{eqnarray}
ds^2&=&A^2(x,y)\,ds^2(\Xsp)
       +B^2(x,y)\,ds^2(\Ysp)\,,
   \label{eq:general-metric}
\end{eqnarray}
%======================================%
where $ds^2(\Xsp)$ denotes the four-dimensional metric 
depending only on the coordinates $x^\mu$ of $\Xsp$, and 
$ds^2(\Ysp)$ denotes the six-dimensional metric depending only 
on the coordinates $y^p$ of  $\Ysp$. 
Hence, the dynamics is essentially limited to the scale factors $A$ and $B$. Concerning the other fields, we adopt the following assumptions
\Eqrsubl{eq:gauge-assumption}{
& &\tau\equiv C_0+i\, e^{-\Phi}=ig_s^{-1}\,,
   \label{eq:gauge-assumption1}
      \\
& &G_3\equiv ig_s^{-1}\,H_3-F_3
  =\frac{1}{3!}\,G_{pqr}\,dy^p\wedge dy^q \wedge dy^r
\,,
   \label{eq:gauge-assumption2}
      \\
& &\ast_{\rm Y}\,G_3=\epsilon\,i\,G_3\quad (\epsilon=\pm 
1)\,,\label{eq:gauge-assumption2-1}
      \\
& &d\ast(B_2 \wedge F_3)=0\,,
   \label{eq:gauge-assumption3}
      \\
& &\tilde{F}_5
  =A^4B^{-4}(1\pm\ast)V_p\,dy^p\wedge\Omega(\Xsp)
  =A^4 B^{-4} V\wedge \Omega(\Xsp)
  \mp \ast_{\rm Y}\,V\,,
  \label{eq:gauge-assumption4}
}
where $g_s$ is a constant representing the string coupling constant, 
and $\ast$ and $\ast_{\rm Y}$ are the Hodge duals with
respect to the ten-dimensional metric $ds^2$ and the six-dimensional 
metric $ds^2(\Ysp)$, respectively.

%T2>Field equations
\subsection{Reduction of the gauge field equations}

Under the assumptions given above, we first reduce the field 
equations other than the Einstein equations to a simple set of 
equations. 

The gauge field equations are given by \cite{Schwarz:1983wa,Schwarz:1983qr}
\Eqrsub{
\Box\tau + i\frac{(\nabla\tau)^2}{\tau_2}
  &=&-\frac{i}{2}G_3\cdot G_3,\\
dG_3&=&0\,,
   \label{eq:gauge-field1}
     \\
\nabla\cdot G_3&\equiv & *d*G_3=-iG_3\cdot\tilde{F}_5
\,,
   \label{eq:gauge-field2}
     \\
\tilde{F}_5&=&dC_4+B_2\,\wedge F_3\,,
   \label{eq:gauge-field3}
     \\
\ast\tilde{F}_5&=&\pm\tilde{F}_5\,.
   \label{eq:gauge-field4}
}
In the present paper, we define the inner product of a $p$-form 
$\omega_p$ and a $q$-form $\chi_q$ ($p\le q$) as
\Eq{
(\omega_p \cdot \chi_q)_{\mu_1\cdots \mu_{q-p}}
 := \frac{1}{p!}\omega^{\nu_1\cdots \nu_p}
   \chi_{\nu_1\cdots \nu_p \mu_1\cdots \mu_{q-p}}.
}
Note that the first equation is automatically satisfied under our 
ansatz.

Under the assumption \eqref{eq:gauge-assumption2}, the equation 
\eqref{eq:gauge-field1} implies that $G_3$ is a closed form 
depending only on the coordinates of $\Ysp$. Then, from the 
self-duality requirement for $G_3$, \eqref{eq:gauge-assumption2-1}, 
it follows that $G_3$ can be expressed in terms of a 2-form 
$\alpha_2$ on $\Ysp$ satisfying $d\ast_{\rm Y} d\alpha_2=0$ as
\begin{eqnarray}
G_3=i\,d\alpha_2+\epsilon\ast_{\rm Y}d\alpha_2.
    \label{eq:ex-g3}
\end{eqnarray}
Therefore, $B_2$ and $F_3$ are expressed as  
\begin{eqnarray}
B_2=g_s\,\alpha_2\,,\qquad
F_3=-\epsilon\,\ast_{\rm Y} d\alpha_2\,.
   \label{eq:bf}
\end{eqnarray}

Next we consider the 5-form $\tilde{F}_5$. First, note that for any  
$q$-form $\omega_q$ on $\Ysp$, the Hodge dual operators $\ast$ and 
$\ast_{\rm Y}$ are related by
\begin{subequations}
\begin{eqnarray}
\ast\omega_q&=& A^4 B^{6-2q} \,\Omega(\Xsp)\,
                    \wedge\,\ast_{\rm Y} \,\omega_q\,,
      \\
\ast\left[\Omega(\Xsp) \wedge \omega_{q}\right]
                  &=&-A^{-4}B^{6-2q}\,\ast_{\rm Y} \omega_{q}\,.
    \label{eq:dual-relation}
\end{eqnarray}
\end{subequations}
Further, the operator $*_{\rm Y}$ satisfies the relations
\begin{eqnarray}
\ast_{\rm Y}(\alpha_2 \cdot d\alpha_2)_Y
   =\alpha_2 \wedge\,\ast_{\rm Y} d\alpha_2\,\quad
\ast_{\rm Y} \ast_{\rm Y} \,\omega_q=(-1)^q\,\omega_q\,,
    \label{eq:dual-y}
\end{eqnarray}
where $(\omega\cdot\chi)_Y$ denote the inner product of forms  
$\omega$ and $\chi$ on $\Ysp$ with respect to the metric 
$ds^2(\Ysp)$. From these relations, we get 
\Eq{
\ast(B_2 \wedge F_3)=g_s \epsilon A^4B^{-4}\,
    \Omega(\Xsp) \wedge \beta_1\,,
   \label{eq:b-f}
}
where $\beta_1$ is the 1-form defined as 
$\beta_1=(\alpha_2\cdot d\alpha_2)_Y$. Hence, the equation 
(\ref{eq:gauge-assumption3}) gives the condition 
\Eq{
d_y(A^4 B^{-4} \beta_1)=0\,.
\label{eq:db-f}
}

Under this condition, the assumption \eqref{eq:gauge-assumption4} on 
$\tilde F_5$ can be rewritten as
\Eq{
(1\pm*)A^4 B^{-4} V\wedge \Omega(\Xsp)
 = d\tilde C_4 
 \pm \epsilon g_s A^4 B^{-4}(1\pm*)\beta_1\wedge \Omega(\Xsp),
}
where $\tilde C_4$ is some 4-form. From this, it follows that $V$ and $\tilde C_4$ can be written in terms of $\beta_1$ and some 1-form $\gamma_1$ as
\Eqrsub{
V&=&\gamma_1\pm \epsilon g_s\,\beta_1\,,
   \label{eq:v-gamma1} \\
d\tilde C_4&=& A^4B^{-4}(1\pm\ast)\gamma_1 \wedge \Omega(\Xsp)
   =A^4B^{-4}\,\gamma_1 \wedge \Omega(\Xsp)
             \mp \ast_{\rm Y} \gamma_1\,.
   \label{eq:5-form}
}
From this and the self duality of $\tilde F_5$, $\gamma_1$ is 
required to satisfy
\Eq{
\pd_x \gamma_1= 0,\quad
d_y(A^4B^{-4}\,\gamma_1)=0,\quad
\hat{D}\cdot\gamma_1=0,
   \label{eq:v-gamma2}
}
where $\hat{D}_p$ is the covariant derivative with respect to the metric $ds^2(\Ysp)$. In particular, $\gamma_1$ is a 1-form on $\Ysp$ independent of $x$.

Finally, utilising the relations
\Eqrsub{
G_3\cdot\tilde{F}_5&=&\pm i\epsilon\,B^{-6}(V\cdot G_3)_Y\,,
      \\
\ast d\ast G_3 &=&A^{-4}B^{-2}\inpare{d_y(A^4)\cdot G_3}_Y \,,
  \label{eq:g3-f5}
}
the remaining field equation \eqref{eq:gauge-field2} can be replaced by the simple equation  
\Eq{
[V\mp \epsilon A^{-4}B^4 d_y(A^4)]\cdot G_3=0\,.
}
Since $G_3$ is self-dual on $\Ysp$, if $G_3\not=0$, this equation is 
equivalent to 
\Eq{
V=\pm \epsilon A^{-4} B^4 d_y(A^4)\,.
}
Since $\beta_1$ and $\gamma_1$ are 1-forms on $\Ysp$, this equation 
and \eqref{eq:v-gamma1} lead to
\Eq{
\partial_\mu (A^{-1}B^4 \partial_p A)=0.
\label{dxV=0}
}
Further, under \eqref{eq:db-f}, the equations \eqref{eq:v-gamma2} are equivalent to 
\Eq{
 \hat D\cdot[A^{-4}B^4 \hat D(A^4)]=g_s (d\alpha_2\cdot d\alpha_2)_Y.
   \label{eq:warp0}
}

To summarise, if we find a 2-form $\alpha_2$ on $\Ysp$ and functions 
$A(x,y)$ and $B(x,y)$ satisfying $d*_{\rm Y} d\alpha_2=0$, 
\eqref{eq:db-f}, \eqref{dxV=0} and \eqref{eq:warp0}, then $B_2$ and 
$F_3$ given by \eqref{eq:bf} and $\tilde F_5$ given by
\Eq{
\tilde F_5=\pm\epsilon (1\pm *)d(A^4)\wedge\Omega(\Xsp)
}
satisfy the field equations other than the Einstein equations. Note 
that this yields the most general solution under our ansatz in the 
case $G_3\not=0$, while in the case $G_3=0$ it may be a special 
solution.

%T2>Einstein Eqs
\subsection{The Einstein equations}

In order to complete the system of equations, we must also consider 
the Einstein equations \cite{Schwarz:1983wa,Schwarz:1983qr}
\Eq{
R_{MN}=\frac{g_s}{4}\,\left[{\rm Re} (G_{MPQ} G^{\ast~PQ}_N)
       -\frac{1}{2}\,G_3\cdot G^{\ast}_3\,g_{MN}\right]
       +\frac{1}{96}\,\tilde{F}_{MP_1\cdots P_4}\,
        \tilde{F}_N^{~\,P_1\cdots P_4}\,.
   \label{eq:einstein}
}
Under our ansatz, these equations become
\Eqrsub{
R_{\mu\nu}&=&-\frac{1}{4} S(x , y)\,g_{\mu\nu}\,,
   \label{eq:einstein-mn}
      \\
R_{\mu p}&=&0\,,
   \label{eq:einstein-mp}
      \\
R_{pq}&=&\frac{1}{4} S(x , y)\,g_{pq}-\frac{V_p V_q}{2h^2}\,,
 \label{eq:einstein-pq}
}
where $S(x , y)$ is
\Eq{
S(x , y)=g_s\,B^{-6}\,(d\alpha_2\cdot d\alpha_2)_Y
          +B^{-10}\,(V\cdot V)_Y\,.
}

First, if we introduce new set of variables $h(x,y)$ and $a(x,y)$ by 
$B=h^{1/4}$ and $A=a h^{-1/4}$, the equation (\ref{eq:einstein-mp}) 
can be written
\Eq{
R_{\mu p}=-\frac{1}{2h} \pd_{\mu} \pd_p h
 +\frac{2}{ah}\pd_p a \pd_\mu h 
 -\frac{3}{a}\pd_\mu \pd_p a
 +\frac{3}{a^2}\pd_\mu a \pd_p a=0.
\label{eq:einstein-mp1}
}
%======================================%
Further, \eqref{dxV=0} reads
\Eq{
-\frac{1}{h}\partial_\mu\partial_p h
  +\frac{4}{ah}\partial_p a \partial_\mu h
 +\frac{4}{a}\partial_\mu\partial_p a
 -\frac{4}{a^2}\partial_\mu a \partial_p a=0.
}
From these two equations, we obtain $\partial_\mu \partial_p \ln 
a=0$. This implies that $a$ can be written $a=a_0(x)a_1(y)$. 
Therefore, we can set $a=1$ by redefining $ds^2(\Xsp), ds^2(\Ysp)$ 
and $h$. Then, these equations reduce to $\pd_\mu \pd_p h=0$. Hence, 
$h$ can be expressed as $h=h_0(x) + h_1(y)$. 
Further, $V$ can be written as $V=\pm \epsilon\,d_y h$, and 
\eqref{eq:warp0} reads $\triangle_Y h_1=-g_s (d\alpha_2\cdot 
d\alpha_2)_Y$. 

Next, taking account of these results, the equation 
(\ref{eq:einstein-mn}) yields
\Eqrsub{
& &R_{\mu\nu}(\Xsp)-h^{-1} D_{\mu} D_{\nu} h_0
   =-\frac{1}{4}h^{-1} \triangle_X  h_0 \,g_{\mu\nu}(\Xsp)\,,
   \label{eq:mn-tf}
     \\
& & R(\Xsp)=0\,,
   \label{eq:mu-tr}
}
where $\triangle_{X}$ is the Laplacian with respect to the metric 
$ds^2({\rm X_4})$\,. If we require that $d_y h\not=0$, these 
equations can be reduced to
\begin{eqnarray}
R_{\mu\nu}(\Xsp)=0,\qquad
D_{\mu} D_{\nu} h_0
       =\frac{1}{4} \triangle_{X} h_0\,g_{\mu\nu}(\Xsp)\,. 
   \label{eq:ricci-h0}
\end{eqnarray}

Finally, taking account of the Poisson equation for $h$ and the 
expression of $V$ in terms of $h$ again, \eqref{eq:einstein-pq} reduces to
\Eq{
 R_{pq}(\Ysp)=\frac{1}{6} R(\Ysp) g_{pq}(\Ysp),\quad
 R(\Ysp)=\frac{3}{2} \triangle_X h_0.
}
These equations are equivalent
\Eq{
R_{pq}(\Ysp) = \lambda g_{pq}(\Ysp),\qquad
\triangle_X h_0=4\lambda,
}
where $\lambda$ is a constant.

%T4>Sol:summary
To summarise, for any  $h$ of the form 
\Eq{
h=h_0(x)+h_1(y)
\label{Sol:general:h}
}
and 2-form $B_2$ on $\Ysp$ satisfying
\Eqrsub{
&& d *_Y dB_2=0,\quad d_y [h^{-2}(B_2\cdot dB_2)_Y]=0,
\label{Sol:general:B2}\\
&& \triangle_Y h_1=-g_s ^{-1} (dB_2\cdot dB_2)_Y,
\label{Sol:general:h1}\\
&& D_\mu D_\nu h_0=\lambda g_{\mu\nu}(X),
\label{Sol:general:h0}
}
the metric 
\Eq{
ds^2(M_{10})=h^{-1/2} ds^2(\Xsp)+ h^{1/2}ds^2(\Ysp)
\label{Sol:general:metric}
}
with
\Eqrsub{
&& R_{\mu\nu}(\Xsp)=0,
\label{Sol:general:R(X)}\\
&& R_{pq}(\Ysp)=\lambda g_{pq}(\Ysp),
\label{Sol:general:R(Y)}
}
and the gauge fields given by
\Eqrsub{
&& H_3=dB_2,\quad F_3=-\epsilon g_s^{-1} *_Y H_3,
\label{Sol:general:G3}\\
&& \tilde F_5=\pm\epsilon(1\pm *) d(h^{-1}) \wedge \Omega(X^4)
\label{Sol:general:F5}
}
yield a solution to the type IIB supergravity. Under our ansatz, 
this is the most general solution in the case $G_3\not=0$ and $d_y 
h\not=0$, while otherwise it may be a special solution.

%T4>h0
Here, note that the conditions \eqref{Sol:general:h0} and 
\eqref{Sol:general:R(X)} strongly restrict the metric $ds^2(\Xsp)$. 
In fact, as is shown in Appendix \ref{Appendix:h0}, $\Xsp$ is required to be locally flat, irrespective of the value of $\lambda$, if $D_\mu 
h_0\not=0$ and $(Dh_0)^2\not=0$. In this case, the general solution 
for $h_0$ is given by
\Eq{
h_0= \frac{\lambda}{2}x^\mu x_\mu + a_\mu x^\mu + b
\label{Sol:general:h0:explicit}
}
in the Minkowki coordinates, where $a_\mu$ and $b$ are constants 
satisfying the condition $a\cdot a\not=0$. On the other hand, if 
$D_\mu h_0\not=0$ and $(Dh_0)^2=0$, there exists a solution only 
when $\lambda=0$. The four-dimensional metric is restricted to the 
form 
\Eq{
ds^2(\Xsp)=\eta_{\mu\nu}dx^\mu dx^\nu + f(x,y,t-z) (dt-dz)^2,
}
where $f(x,y,t-z)$ is an arbitrary solution to
\Eq{
(\partial_x^2+\partial_y^2)f=0.
}
$h_0$ is expressed in these coordinates as 
\Eq{
h_0= a (t-z) + b,
}
where $a$ and $b$ are arbitrary constants. Note that when $f$ is 
linear with respect to $x$ and $y$, $ds^2(\Xsp)$ is really a flat 
metric and $f$ can be set to zero by a coordinate transformation. In 
this case, $h_0$ is given by \eqref{Sol:general:h0:explicit} with 
$\lambda=0$ and $a\cdot a=0$.

Finally, note that in the case $G_3=0$, the equation 
\eqref{Sol:general:B2} becomes trivial, and \eqref{Sol:general:h0} 
reduces to $\triangle_Y h_1=0$. In the special case of $\lambda=0$, 
the corresponding dynamical solution with 
\eqref{Sol:general:h0:explicit} is identical to the solution found 
by Gibbons, L\"u and Pope \cite{Gibbons:2005rt}. 

%T1>TD KS sols
%======================================%
%<<<<<<<<<<<<< SECTION 3 >>>>>>>>>>>>>>%
%======================================%

%\section{Time-dependent Fractional D3-brane solutions}
\section{Application to the conifold-type compactifications}
\label{sec:conifold}

In this section, we apply the general formulation developed in the 
previous section to the flux compactification on conifold-type 
Calabi-Yau spaces in order to find time-dependent generalisations of 
the Klebanov-Strassler solutions
\cite{Klebanov:2000hb}.  

In all cases, we look for solutions whose ten-dimensional metrics 
have the form%
\Eq{
ds^2= h^{-1/2}(x,r) ds^2(E^{3,1})+ h^{1/2}(x,r) ds^2(\Ysp),
}
where $\Xsp=E^{3,1}$ is the four-dimensional Minkowski spacetime, 
$\Ysp$ is a six-dimensional Calabi-Yau space, and $r$ is a radial 
coordinate of $\Ysp$. Further, we only consider Calabi-Yau spaces 
whose level surface with respect to $r$ approaches the Einstein 
space $\Tsp$ at large $r$ (up to a scale factor), whose metric is given 
by \cite{Candelas:1989js}\,   
\Eq{
ds^2(\Tsp)=\frac{1}{9} 
       \left(d\psi + \sum_{i=1}^{2} \cos \theta_i\,d\phi_i\right)^2
     + \frac{1}{6} \sum_{i=1}^{2}
       \left(d\theta_i^2 + \sin^2\theta_i\,d\phi^2_i\right),
    \label{eq:t11}
}
where the range of the angular coordinates $\theta_i$, $\phi_i$\, 
and $\psi$ are $0\le \theta_i < \pi$, $0\le\phi_i<2\pi$ and $0\le 
\psi< 4\pi$, respectively. 

Throughout this section, we use the following orthogonal basis
\cite{Klebanov:2000hb,Minasian:1999tt}:
\begin{eqnarray}
& &g^1=\frac{1}{\sqrt{2}} (e^1-e^3)\,, \quad
   g^2=\frac{1}{\sqrt{2}} (e^2-e^4)\,, \quad
   g^3=\frac{1}{\sqrt{2}} (e^1+e^3)\,,
      \nn\\
& &g^4=\frac{1}{\sqrt{2}} (e^2+e^4)\,, \quad
   g^5=e^5\,,
    \label{eq:basis-g}
\end{eqnarray}
where 
\begin{eqnarray}
& &e^1\equiv -\sin\theta_1\,d\phi_1\,,\quad
   e^2\equiv d\theta_1\,, \quad
   e^3\equiv \cos \psi\,\sin \theta_2\, d\phi_2\,
          - \sin\psi\,d\theta_2\,,\nn\\
& &e^4\equiv \sin \psi\,\sin \theta_2\,d\phi_2\,
          + \cos \psi\,d\theta_2\,,\quad
   e^5\equiv d\psi + \cos \theta_1\,d\phi_1\,
          + \cos\theta_2\,d\phi_2\,.
    \label{eq:basis-t11}
\end{eqnarray}
The line element of $\Tsp$ is expressed in terms of this basis as
\Eq{
ds^2(\Tsp)=\frac{1}{9} (g^5)^2
        +\frac{1}{6} \sum_{i=1}^{4}  (g^i)^2\,.
\label{eq:bt11}
}
%

%T2>TD conifold solution
\subsection{Time-dependent conifold solution}
\label{subsec:TDCS}

First, we consider the case in which $\Ysp$ is a simple conifold over $\Tsp$: 
\Eq{
ds^2(\Ysp)=dr^2+r^2 ds^2(\Tsp).
\label{eq:calabi-metric}
}

Let $B_2$ be a 2-form on $\Ysp$ of the form \cite{Klebanov:2000hb}%
\Eq{
B_2 = g_s\,f(r) (g^1 \wedge g^2 
         +g^3 \wedge g^4)
    = g_s\,f(r) 
      \left[\Omega({\rm S}_1^2)-\Omega({\rm S}_2^2)\right],
  \label{eq:b2}
}
where
\Eq{
\Omega({\rm S}_1^2)=\sin\theta_1\,d\theta_1 \wedge d\phi_1\,,\quad
\Omega({\rm S}_2^2)=\sin\theta_2\,d\theta_2 \wedge d\phi_2\,.
}
Then, from \eqref{Sol:general:G3}, $H_3$ and $F_3$ are given by
\Eqrsub{
H_3&=&dB_2=g_s\,f'(r) dr \wedge 
    \left[\Omega({\rm S}_1^2)-\Omega({\rm S}_2^2)\right],
  \label{eq:h3}
      \\
F_3&=&-\epsilon g_s^{-1}\,\ast_{\rm Y} H_3
    =-\epsilon \frac{1}{3} r f'(r) d\psi \wedge 
     \left[\Omega({\rm S}_1^2)-\Omega({\rm S}_2^2)\right].
  \label{eq:f3}
}
Hence, the first equation of \eqref{Sol:general:B2} gives 
$rf'=M=\const$:
\Eq{
f(r)=M \ln\left(\frac{r}{r_0}\right),
    \label{eq:f}
}
where $r_0$ is a constant.  Further, the second equation of 
\eqref{Sol:general:B2} is trivially satisfied, because $(B_2\cdot 
H_3)_Y$ is a constant multiple of $ff'dr$ and $h$ depends only on 
$r$ for fixed $x$-coordinates.  

The remaining non-trivial equation \eqref{Sol:general:h1} reduces to
\Eq{
\hat{\triangle}_{\rm Y_6} h=\frac{1}{r^5} (r^5 h')'
      =-g_s^{-1} (H_3\cdot H_3)_Y
      =-\frac{72g_s M^2}{r^6}\,.
}
Thus, taking account of \eqref{Sol:general:h} and  
\eqref{Sol:general:h0:explicit}, the general solution for $h$ under 
our ansatz is given by %
\Eq{
h(x , r) =h_0(x)+\frac{36 g_s\,M^2}{r^4}
   \left[\ln\left(\frac{r}{r_0}\right)+\frac{1}{4}\right]
   +\frac{C}{r^4}\,,
    \label{eq:frac-sol}
}
where $C$ is a constant, and $h_0(x)$ is a linear function of 
$x^\mu$.

Now, let us briefly discuss some characteristic features of this 
solution (cf. Ref. \cite{Gibbons:2005rt}). First, in the region 
where $r\gg |C|, g_s M^2 |\log(r/r_0)+1/4|$, the corresponding 
spacetime metric depends only on $h_0(x)$, and there appears 
curvature singularity at the hypersurface where $h_0(x)$ vanishes, 
if $D h_0\not=0$. For example, for $h_0(t)=-p\,t+q$ ($p>0$), the 
four-dimensional metric $h^{-1/2}ds^2(E^{3,1})$ represents an 
expanding universe for $t<q/p$, which ends at the big-lip 
singularity at $t=-q/p$\,. This expansion is associated with  
contraction of the internal space. Hence, if we follow the standard 
prescription in which the effective scale factor of the 
four-dimensional universe is given by $A B^3=h^{1/2}$, this solution 
is interpreted as representing a contracting universe. The converse 
situation arises for $p<0$. However, this interpretation based on 
the effective action may not be valid in the case in which moduli 
are not stabilised, because changes in moduli produce changes in 
fundamental coupling constants, which affect the spectra of atoms 
for example. A correct physical interpretation should be obtained 
only by taking account of such effects on observations.

In contrast to this large $r$ region, the time dependence of $h$ in 
the small $r$ region is negligible compared with the terms produced 
by flux, provided that $h>0$. Hence, in this region, the scale 
modulus is practically stabilised for a long time. This feature may 
be used as a moduli stabilisation mechanism in the context of the 
braneworld model \cite{Randall:1999ee,Randall:1999vf}, 
in particular for a similar solution in the 
deformed conifold compactification discussed later.

%T2>Resolve conifold
\subsection{Resolved-conifold compactification}

Next, let us consider time-dependent solutions for compactification 
on the resolved conifold, whose metric is given by
\cite{PandoZayas:2000sq,Papadopoulos:2000gj,PandoZayas:2001iw}
\Eq{
ds^2(\Ysp)=\kappa^{-1}(r) dr^2+\frac{1}{9} \kappa(r) r^2\,(e^5)^2
         +\frac{1}{6} r^2 ds^2({\rm S}_1^2)
         +\frac{1}{6} (r^2+6a^2) ds^2({\rm S}_2^2)\,,
}
where 
\Eq{
\kappa(r)=\frac{r^2+9a^2}{r^2+6a^2}\,,
}
and $ds^2({\rm S}_1^2)$ and $ds^2({\rm S}_2^2)$ denote the line 
elements of the spheres ${\rm S}_1^2$ and ${\rm S}_2^2$, 
respectively. 

For this internal space, the choice of $B_2$ of the form
\cite{PandoZayas:2000sq, PandoZayas:2001iw}
\Eq{
B_2 = g_s \left[f_1(r) \Omega({\rm S}_1^2)
      +f_2(r) \Omega({\rm S}_2^2)\right]
   \label{eq:r-b2}
}
yields
\Eqrsub{
H_3&=&dB_2=g_s\,dr \wedge \left[f'_1(r) \Omega({\rm S}_1^2)
         +f'_2(r) \Omega({\rm S}_2^2)\right],
   \label{eq:r-h3}
       \\
F_3&=&\epsilon \frac{1}{3} g^5 \wedge \left[\frac{r^2+6a^2}{r} 
      f'_1(r) \Omega({\rm S}_2^2)+\frac{r^3}{r^2+6a^2} 
      f'_2(r) \Omega({\rm S}_1^2)\right],
   \label{eq:r-f3}
}
where the prime denotes differentiation with respect to $r$. 
>From the first equation of \eqref{Sol:general:B2}, $dF_3=0$,  the 
functions $f_1(r)$ and $f_2(r)$ obey
\Eq{
f'_1(r)=\frac{M r}{r^2+9a^2}\,,\quad
f'_2(r)=-\frac{M (r^2+6a^2)^2}{r^3(r^2+9a^2)}\,.
    \label{eq:df1-df2}
}
The general solution to these equations are 
\Eqrsub{
f_1(r)&=&C_1+\frac{1}{2} \ln(r^2+9a^2)\,,\\
f_2(r)&=&C_2+\frac{2a^2}{r^2}
        -\frac{1}{18} \ln\left[r^{16} 
    (r^2+9a^2)\right],\label{eq:f1-f2}
}
where $C_1$ and $C_2$ are constants. The equation 
\eqref{Sol:general:h1} now reads
\Eq{
\left[r^3 (r^2+9a^2) h'\right]'=-36 g_s M^2 
       \frac{r^4+(r^2+6a^2)^2}{r^3 (r^2+9a^2)}\,,
    \label{eq:warp-equation}
}
where we have used 
\begin{eqnarray}
\beta_1&=&36 g_s \frac{dr}{r^3 (r^2+9a^2)} (f_1-f_2)\,,
      \\
H_3\cdot H_3&=&36 g_s^2 M^2  
       \frac{r^4+(r^2+6a^2)^2}{r^6 (r^2+6a^2) (r^2+9a^2)}\,.
    \label{eq:b1-h3}
\end{eqnarray}
The second of \eqref{Sol:general:B2} is again trivially satisfied. 
Therefore, we obtain 
\Eq{
h' = \frac{4g_s M^2}{r^3 (r^2+9a^2)}
    \left(\frac{18a^2}{r^2}-\ln\left[r^8 (r^2+9a^2)^5\right]\right)
    +\frac{C}{r^3 (r^2+9a^2)}\,,
   \label{eq:resolved-equations}
}
where $C$ is a constant.

For large $r$ ($r\gg 3a$)\,, we reproduce the solution 
(\ref{eq:frac-sol}) after integrating \eqref{eq:resolved-equations} 
over $r$, taking account of \eqref{Sol:general:h} and 
\eqref{Sol:general:h0:explicit}. On the other hand, for small $r$ 
($r\ll 3a$)\,, the solution is approximated 
by \cite{PandoZayas:2000sq}
\Eq{
h(x,r)=h_0(x)-\frac{C}{18a^2 r^2}-\frac{2g_s M^2}{r^4}\,,\label{eq:r
esolved-field}
}
where $h_0(x)$ is a linear function of $x$\,.

%T2>Deformed conifold
\subsection{Deformed-conifold compactification}
  \label{subsec:deformed}
  
Finally, we show that the deformed-conifold solution also has a 
dynamical generalisation. 

The deformed-conifold metric can be written as 
\cite{Klebanov:2000hb,Minasian:1999tt}
\begin{eqnarray}
ds^2_6(\Ysp)&=&\frac{1}{2} \sigma^{2} K(\tau) 
   \left[\frac{1}{3K^3(\tau)} \left\{d\tau^2+(g^5)^2\right\}
   +\sinh^2\left(\frac{\tau}{2}\right) \left\{(g^1)^2+(g^2)^2\right\}
   \right.
       \nn\\
   & &\qquad
   \left.+\cosh^2\left(\frac{\tau}{2}\right) 
          \left\{(g^3)^2+(g^4)^2\right\} \right],
    \label{eq:deformed-metric}
\end{eqnarray}
where $\sigma$ is a constant, and 
\begin{eqnarray}
K(\tau)=\frac{\left[\sinh(2\tau)-2\tau\right]^{1/3}}
           {2^{1/3} \sinh(\tau)}\,.
\end{eqnarray}
The radial coordinate $r$ is related to $\tau$ by
\Eq{
\frac{r}{\sigma}=6^{-1/2}\,\sigma\,\int_0^\tau \frac{du}{K(u)}.
}

For this Calabi-Yau space, let us assume that $B_2$ takes the form 
\cite{Klebanov:2000hb}
\Eq{
B_2= g_s \left[k_1(\tau) 
      g^1 \wedge g^2+k_2(\tau) 
      g^3 \wedge g^4\right].
   \label{eq:d-b2}
}
Then, from \eqref{Sol:general:G3}, $H_3$ and $F_3$ are expressed as
\Eqrsub{
H_3&=&dB_2=g_s \Big[d\tau \wedge \left(k'_1 
      g^1 \wedge g^2+k'_2 
      g^3 \wedge g^4\right)
       \nn\\
     & &\qquad\qquad 
      -\frac{1}{2} (k_1-k_2) g^5 \wedge (g^1 \wedge g^3+
      g^2 \wedge g^4)\Big],
   \label{eq:d-h3}
        \\
F_3&=&M \left[
      -g^5 \wedge \left\{\frac{k_1'}{\tanh^2(\tau/2)} g^3 \wedge g^4
      +k_2' \tanh^2(\tau/2) \wedge g^1 \wedge g^2\right\}\right.
       \nn\\
    & &\left. +\frac{1}{2}(k_1-k_2) d\tau \wedge 
       (g^1 \wedge g^3+g^2 \wedge g^4)\right],
   \label{eq:d-f3}
}
where the prime denotes the differentiation with respect to $\tau$, 
and we have used the relation 
\Eq{
d(g^1 \wedge g^2)=-d(g^3 \wedge g^4)
       =-\frac{1}{2} g^5 \wedge (g^1 \wedge g^3+g^2 \wedge g^4)\,.
}

Now, utilising the relations
\begin{eqnarray}
& &d(g^5 \wedge g^3 \wedge g^4)
    =d(g^5 \wedge g^1 \wedge g^2)=0\,,
       \\
& &d(g^1 \wedge g^3+g^2 \wedge g^4)
    =g^5 \wedge (g^1 \wedge g^2-g^3 \wedge g^4)\,,
\end{eqnarray}
the first of \eqref{Sol:general:B2} reduces to
\Eq{
\left(\frac{k'_1}{\tanh^2(\tau/2)}\right)'=\frac{k_1-k_2}{2}\,,\quad
\left[k'_2 \tanh^2(\tau/2)\right]'=-\frac{k_1-k_2}{2}\,.
    \label{eq:d3=0}
}
These equations are equivalent to
\Eq{
k'_1=\alpha (1-F) \tanh^2(\tau/2)\,,\quad
k'_2=\alpha F \coth^2(\tau/2)\,,
   \label{eq:deform-dif-k1k2}
}
where $\alpha$ is a constant and $F(\tau)$ obeys the differential 
equation
\Eq{
F''=\frac{1}{2} \left[F \coth^2\left(\frac{\tau}{2}\right)
    +(F-1) \tanh^2\left(\frac{\tau}{2}\right)\right]\,.
   \label{eq:deform-dif-F}
}
Since $(B_2\cdot H_3)_Y$ is a constant multiple of $(k_1k_1' + 
k_2k_2')d\tau$ and $h$ depends only on $\tau$ and $x$, the second 
equation of \eqref{Sol:general:B2} is automatically satisfied as in 
the previous cases. Hence, from the general arguments in 
\S\ref{sec:KS-solution}, for each solution to 
\eqref{eq:deform-dif-F}, we obtain a dynamical solution of the form 
$h=h_0(x)+h_1(y)$, where $h_0(x)$ is a linear function of $x^\mu$ 
and $h_1$ is a solution to \eqref{Sol:general:h1}.

For example, we require that $h$ is regular at $\tau=0$ and 
approaches a constant at $\tau=\infty$, $F$ and $h$ are determined as 
\cite{Klebanov:2000hb}
\Eqrsub{
&& F=\frac{\sinh\tau -\tau}{2\sinh\tau},\\
&& h(x , \tau)=h_0(x)
   +\alpha \frac{2^{2/3}}{4} \int^{\infty}_{\tau}\,du\,
    \frac{u \coth u-1}{\sinh^2 u}\left\{\sinh(2u)-2u\right\}^{1/3}\,.
   \label{eq:large-h2}
}
At large $r$, this solution behaves as
\Eq{
h \sim h_0(x) + \frac{81 \alpha\sigma^4}{2r^4}\ln\frac{r}{\sigma}.
}
This behavior is the same as that of the conifold solution 
\eqref{eq:frac-sol}. From this we find that $\alpha\sigma^4$ 
corresponds to $g_s M^2$ representing the intensity of the $G_3$ 
flux. Hence, for a large $G_3$ flux, this solution provides a 
regular solution with a large warp factor. As discussed in 
\S\ref{subsec:TDCS}, this large warp factor stabilises the scale 
modulus for cosmological solutions with $h_0=-p\,t +q$. 

For reference, we gave an explicit expression for the general 
solution, which is in general singular at $r=0$ or at $r=\infty$, in 
Appendix \ref{Appendix:deformed}.

%T1>SUSY breaking
%======================================%
%<<<<<<<<<<<<< SECTION 4 >>>>>>>>>>>>>>%
%======================================%

\section{Supersymmetry breaking}
\label{sec:susy}

In this section, we examine whether supersymmetry is preserved or 
not by the dynamical instability of the scale modulus. For 
simplicity, we only consider the case of no 3-form flux, 
$B_2=C_2=C_0=\Phi=0$ and assume that the ten-dimensional metric has 
the form
\Eqr{
ds^2&=&g_{MN} dx^M dx^N
      \nn\\
    &=&h^{-1/2}(x , r) ds^2(E^{3,1})
      +h^{1/2}(x , r) [dr^2+r^2 ds^2(\Zsp)]\,.
   \label{eq:susy-metric}
}

%T2>SUSY trf
\subsection{Supersymmetry transformation}

In the ten-dimensional type IIB supergravity with 
$B_2=C_2=C_0=\Phi=0$, the local supersymmetry transformation of the 
spinor fields (gravitino $\psi_M$ and dilatino $\lambda$) is given 
by \cite{Schwarz:1983wa,Schwarz:1983qr,Kehagias:1998gn,  
Howe:1983sr,Duff:1991pe}
\Eqrsub{
\delta  \lambda &=& 0\,,
  \label{eq:susy-transform-psi2}
     \\
\delta \psi_{M}&=&\bar{\nabla}_M \epsilon\,, 
  \label{eq:susy-transform-lambda2}
}
where $\epsilon$ is a ten-dimensional complex Weyl spinor satisfying 
the chirality condition $\Gamma^{11}\epsilon=\pm1$, and the 
covariant derivative $\bar{\nabla}_M$ is given by 
\Eq{
\bar{\nabla}_M = \nabla_M 
     + \frac{i}{16\cdot 5!} \slash{F}_5\,\Gamma_M
   \label{eq:derivative2}
}
in terms of the ten-dimensional $\gamma$-matrices $\Gamma^M$  
satisfying 
\Eq{
\Gamma^M \Gamma^N + \Gamma^N \Gamma^M=2g^{MN}.
}
The bosonic fields are automatically invariant under local 
supersymmetric transformations because we are considering solutions 
with vanishing spinor fields.

For the metric \eqref{eq:susy-metric}, it is convenient to introduce 
$\gamma^\mu$($\mu=0,\cdots,3$), $\gamma^r$ and 
$\gamma^l$($l=5,\cdots,9$) by
\Eq{
\Gamma^{\mu}=h^{1/4} \gamma^{\mu}\,,\quad
\Gamma^{r}=h^{-1/4} \gamma^{r}\,,\quad
\Gamma^{l}=\frac{1}{r h^{1/4}} \gamma^{l}\,. 
   \label{eq:large-gamma}
}
Then, $\gamma^{\mu}$ give the $\SO(3,1)$ $\gamma$-matrices, 
$\gamma^l$ provide the $\gamma$-matrices of $\Zsp$, and 
$(\gamma^r)^2=1$. We also define $\gamma_{(4)}$ and $\gamma_{(10)}$ 
by
\Eq{
\gamma_{(4)}=-i\,\gamma^0\,\gamma^1\,\gamma^2\,\gamma^3\,,
     \quad
\gamma_{(6)}= \gamma_{(4)}\Gamma^{11},
   \label{eq:g4-g6}
}
so that $(\gamma_{(4)})^2=(\gamma_{(6)})^2=1$\,. 

In terms of these $\gamma$-matrices, the supersymmetry 
transformation in the background with the metric 
(\ref{eq:susy-metric}) is expressed as
\Eqrsubl{eq:killing}{
\bar{\nabla}_{\mu}\,\epsilon&=&
       \left[\prefix{X}{\nabla}_{\mu}
       +\frac{\pd_\nu  h}{8 h} {\gamma^{\nu}}_{\mu}
       -\frac{h'}{8 h^{3/2}} \gamma_{\mu}\,\gamma^r
%       \,\{\gamma_{(4)}(1\pm\Gamma^{11})-2\}
       \right] \epsilon\,,
    \label{eq:killing-m}
       \\
\bar{\nabla}_r \epsilon&=&
      \left[\partial_r+\frac{h'}{8 h} \gamma_{(4)} 
       -\frac{\pd_\mu  h}{8 h^{1/2}} \gamma^{\mu}\,\gamma^r\right]
         \epsilon\,,
    \label{eq:killing-r}
        \\
\bar{\nabla}_{l} \epsilon&=&\left[\prefix{Z}{\nabla}_l  
       +\frac{1}{2} \gamma_l\,\gamma^r
       -\frac{r \pd_\mu  h}{8 h^{1/2}} \gamma^{\mu}\,\gamma_l
       -\frac{r h'}{8h} \gamma^r\,\gamma_l
%         \{\gamma_{(4)}(1\pm\Gamma^{11})-2\}
       \right] \epsilon\,,  
    \label{eq:killing-l}
}
where the prime denotes differentiation with respect to $r$, and 
${}^{X}\nabla_{\mu}$ and ${}^{Z}\nabla_l$ are the covariant 
derivatives with respect to the metrics, $ds^2(E^{3,1})$ and 
$ds^2({\rm Y}_5)$, respectively. The number of unbroken 
supersymmetries is determined by the number of covariantly constant 
(or Killing) spinor $\epsilon$ for which the right-hand sides of 
\eqref{eq:killing} vanish.  

%T2>Number of SUSY
\subsection{Consistency condition for the Killing spinor}

Each generator $\epsilon$ of an unbroken supersymmetry should 
satisfy the consistency (integrability) condition
\Eq{
[\bar{\nabla}_M ,~ \bar{\nabla}_N] \epsilon =0\,.
   \label{eq:integ-condition}
}
By using the relation 
\Eq{
\nabla_M \Gamma^N=0\,,\quad
\left[\nabla_M ,~ \nabla_N\right]
      =\frac{1}{4} R_{MNPQ} \Gamma^{PQ}\,,
   \label{eq:sub-condition}
}
the commutator of the covariant derivatives in the consistency 
condition (\ref{eq:integ-condition}) can be in general 
written
\begin{eqnarray}
[\bar{\nabla}_M ,~ \bar{\nabla}_{N}]
   &=&\frac{1}{4} R_{MNPQ} \Gamma^{PQ}
      +\frac{2i}{16\cdot 5!} \nabla_{[M}\,\slash{F}_5\,\Gamma_{N]}
      -\frac{1}{(16\cdot 5!)^2} 
      [\slash{F}_5\,\Gamma_M ,~ \slash{F}_5\,\Gamma_N]
      \nn\\
   &=&\frac{1}{4} R_{MNAB} \Gamma^{AB}
     -\frac{i}{96} \nabla_{P_1} F_{P_2 P_3 P_4 M N} 
      \Gamma^{P_1 \cdots  P_4} (1\pm\Gamma^{11})
      \nn\\
    & & -\frac{1}{64\cdot 4!} {F^{Q_1 \cdots Q_4}}_P 
       F_{Q_1 \cdots Q_4 [M} 
         \Gamma^P_{~ N]} (1\pm\Gamma^{11})\,.
   \label{eq:integ-condition1}
\end{eqnarray}

With the help of this consistency condition, let us examine how many 
supersymmetries exist. To begin with, for comparison, we recall the 
results for the well-studied case of the static 
background \cite{Kehagias:1998gn,Duff:1991pe,Romans:1984an,Kim:1985ez}.
 First, for the case in which  $h=\const $ or $h=C/r^4$, 
the only non-trivial consistency condition is given by
\Eq{
[\bar{\nabla}_{l} ,~ \bar{\nabla}_{m}] \epsilon
  =\frac{1}{4} C_{lmpq}(\Zsp) \gamma^{pq} \epsilon\,,
   \label{eq:consntant-consistency}
}
where $C_{lmpq}$ is the Weyl tensor of the $\Zsp$ space. Hence, the 
number of supersymmetries is determined by the number of solutions 
to the spinor equation
%============< EQUATION >==============%
%
\begin{eqnarray}
\bar{\nabla}_l \epsilon_0
   =\left({}^{\rm Y_5}\nabla_l \pm
    \frac{1}{2} \gamma_l \right) \epsilon_0=0\,.
    \label{eq:consntant-killing}
\end{eqnarray}
%======================================%
In particular, for the ten-dimensional Minkowski 
spacetime \cite{Schwarz:1983qr} and for $\AdS^5\times 
{\rm S}^5$ \cite{Kim:1985ez}, the background has the full supersymmetry.

Next, we consider the static conifold background with 
$h=h_0+C/r^4$($h_0 C\not=0$) \cite{Kehagias:1998gn}. For this 
background, the $[\mu,r]$-component of the consistency condition 
reads
\Eq{
0=[\bar{\nabla}_{\mu} ,~ \bar{\nabla}_{r}] \epsilon
  =2 h^{-1/4} (h^{-1/4})^{''}\,\gamma_{\mu}\,\gamma^{r}\,
   \left(\gamma_{(4)}-1\right)\epsilon\,.
    \label{eq:static-consistency}
}
Hence, $\epsilon$ should satisfy 
\Eq{
\left(\gamma_{(4)}-1\right)\epsilon=0\,.
    \label{eq:static-killing}
}
We can show that this condition and \eqref{eq:consntant-consistency} 
are the only non-trivial consistency conditions. Hence, one half of 
the supersymmetries in the previous case are broken 
in this conifold background \cite{Kehagias:1998gn, Romans:1984an}.

Now, let us consider the background with $\pd_{\mu} h\ne 0$\.. In 
this case, from the $[\mu,\nu]$-components of the consistency 
condition, we obtain
\Eq{
0=b^{\mu} c^{\nu} [\bar{\nabla}_{\mu} ,~ \bar{\nabla}_{\nu}] 
  \epsilon
 =-\frac{(D h_0)^2}{32 h^{2}} b_{\mu}\gamma^{\mu} c_{\nu} 
 \gamma^{\nu} \epsilon\,,
   \label{eq:susy-condition}
}
where $b_{\mu}$ and $c_{\nu}$ are linearly independent vectors 
satisfying the conditions $b^{\mu} \pd_{\mu} h=c^{\mu} \pd_{\mu} 
h=0$\,.  From this, it follows that if $a_\mu=\partial_\nu h_0$ is 
not a null vector, there exists no non-trivial solution to the 
consistency condition, and the supersymmetry is completely broken. 
In contrast, when $a_\mu$ is a null vector, we find that the 
consistency condition is equivalent to
\Eq{
\gamma_{(4)} \epsilon=\epsilon ,\quad
(D_{\mu} h) \gamma^{\mu} \epsilon=0\,,\quad
\frac{1}{2} C_{lmpq}({\rm Y}_5) \gamma^{pq} \epsilon=0\,.
\label{eq:susy-assumption}
}
%======================================%
Hence, a quarter of the supersymmetries in the case of $h_0 C=0$ are 
preserved.

Finally, we comment on the degree of supersymmetry breaking for the 
dynamical background. One natural measure for that is obtained from 
\eqref{eq:susy-condition}. It is the mass scale corresponding to 
$(Dh_0)^2/h^2$. If we consider the induced effective mass for the 
spinor field, we obtain a similar mass scale. This mass scale 
diverges at the naked singularity where $h$ vanishes. Hence, for the 
cosmological situation $h=-p\,t + q$ argued in the previous section, 
the degree of supersymmetry breaking increases as the universe 
approaches the big-lip singularity. In contrast, in the region with 
a large warp factor, the supersymmetry breaking becomes negligible.

%T1>Discussion
%======================================%
%<<<<<<<<<<<<< SECTION 5 >>>>>>>>>>>>>>%
%======================================%

\section{Discussion}
\label{sec:discussion}
In the present paper, we have studied the dynamical stability in the 
moduli sector of supersymmetric solutions for the conifold-type 
warped compactification of the ten-dimensional type IIB 
supergravity, by looking for extensions of supersymmetric solutions 
to those that depends on the four-dimensional coordinates. We have 
found that for many of the well-know solutions compactified on a 
conifold, resolved conifold or deformed conifold, such extensions 
exist and exhibit a dynamical instability. Further, this instability 
is associated with supersymmetry breaking. This feature is expected 
to be shared  by a quite wide class of supersymmetric solutions 
beyond the examples considered in the present paper, because the 
result has been obtained by analysing the general structure of 
solutions for warped compactification with flux of the type IIB 
supergravity under ansatz that is natural to include supersymmetric 
solutions as a special case.

The dynamical solutions found in the present paper can be always 
obtained by replacing the constant modulus $h_0$ in the warp factor 
$h=h_0+h_1(y)$ for supersymmetric solutions by a linear function 
$h_0(x)$ of the four-dimensional coordinates $x^\mu$. Since $h_0$ 
corresponds to the scaling degree of freedom of the internal space, 
this implies that the dynamical instability occurs in the K\"ahler 
modulus representing the scale of the internal space. This type of 
instability in the moduli sector itself is not surprising, because 
constant moduli have flat potentials in effective four-dimensional 
theories. 
In fact, in the large $r$ region in which $h$ becomes independent of $r$, it is expected that the behavior of the solution is well described by an effective four-dimensional theory. Hence, the instability found in the present paper corresponds to a decompactifying run-away solution in an effective theory.
However, it is not expected generally in effective 
theories that such instabilities give rise to significant position-sensitive time dependence in 
the structures of the four-dimensional spacetime and of 
the internal space. 
It is because most effective theories do not 
take account of the warped structure \cite{Kachru:2003aw,Giddings:2001yu}. 
In fact, we have found that 
the degree of instability significantly depends on the position in 
the internal space. In the conifold-type examples considered in the 
present paper, the instability is most enhanced at infinity, while 
near the conifold singularity or in the region with a large warp 
factor for the deformed-conifold case, the instability is strongly 
suppressed. Thus, the moduli stability is closely connected with the 
large warp factor, which has been used to resolve the hierarchy 
problem in the context of the flux
compactification \cite{Giddings:2001yu}.  This feature 
may also play an important role in constructing realistic universe 
model in the KKLT scheme or in the braneworld scheme.

In this connection, we would like to comment on some subtle points. 
First, the instability we have found is a global mode on an open 
internal space. Hence, one may suspect that such an instability does 
not really occur in a model with a compact internal space. 
In particular, in a model in which the scale modulus is stabilised 
by quantum effects in the effective four-dimensional theory, the 
instability may be able to grow only when the effective kinetic 
energy $\dot h^2/h^2$ of the modulus exceeds the height of the 
potential barrier, taking account of the correspondence between the 
ten-dimensional theory and the effective theory mentioned above. 
However, it is quite difficult to make clear this point by an 
explicit analysis because there exists no warped ten-dimensional 
model that takes account of quantum effects and their backreaction 
on the geometry. Actually, there exists at present no warped 
ten-dimensional model with smooth compact internal space in which 
the backreactions of flux and negative charges of orientifold 
planes are properly taken into account, because such negative charges 
produce naked singularities. Apart from this global problem, 
there is also a possibility that a similar instability occurs 
locally. Such a local instability may grow even in a model with 
quantum moduli stabilisation if the spatial scales of the 
instability are smaller than the length scale corresponding to the 
stabilisation energy scale. 
It will be interesting to see whether such a local instability 
exists by a linear perturbation analysis (cf. Ref. 
\cite{Kang:2004hm}). 

Another subtle point is that we have found instability 
only in the scale modulus. For example, in the 
Gibbons-L\"u-Pope solution, the warp factor $h$ can have a large 
number of constant moduli corresponding to the positions of $D3$ 
branes in addition to the scale modulus. Such a solution is 
contained in the class of solutions analysed in the present paper, 
but no instability has been found in these additional moduli. One 
possible reason for this is that the ansatz concerning the structure 
of the ten-dimensional metric is too restrictive. A linear 
perturbation analysis may also be useful in clarifying this point.

Finally, we would like to point out that the degree of supersymmetry 
breaking is also closely related to the warp factor, which can be 
interpreted as the cosmic scale factor in the cosmological context. 
Hence, the cosmic expansion, the hierarchy and the supersymmetry 
breaking are tightly connected. Although the examples considered in 
the present paper do not provide realistic cosmological models, this 
feature may be utilised to solve the hierarchy problem and the 
supersymmetry breaking problem in a realistic higher-dimensional 
cosmological model.

%T1>Acknowledgments
\section*{Acknowledgments}
      
The authors would like to thank Shigeki Sugimoto for valuable 
discussions. This work is supported by the JSPS grant No. 15540267 
(H.~K.) and by the Yukawa fellowship (K.~U.). 

%T1>Appendix
\section*{Appendix}
\appendix

%======================================%
%<<<<<<<<<<<<< APPENDIX >>>>>>>>>>>>>>>%
%======================================%
%T2>Appendix: h0
\section{Solutions for $h_0$ and $\Xsp$}
\label{Appendix:h0}

In this appendix, we find a general solution for $h_0$ and 
$ds^2(\Xsp)$ to \eqref{Sol:general:h0} and 
\eqref{Sol:general:R(X)}.

First, we consider the case in which $D_\mu h_0\not=0 $ and 
$(Dh_0)^2\not=0$. In this case, in terms of the synchronous 
coordinates with respect $h_0=t$, $ds^2(\Xsp)$ can be written
\Eq{
ds^2= N(t,z) dt^2 + q_{ij}(t,z)dz^i dz^j.
}
In this coordinate system, the equation
\Eq{
D_\mu D_\nu h_0=-\Gamma^t_{\mu\nu}=\lambda g_{\mu\nu}
}
is equivalent to 
\Eq{
N^{-1}\partial_t N=-2\lambda N,\quad
\partial_i N=0,\quad
\frac{1}{2N}\partial_t q_{ij}= \lambda .
}
A general solution to this equation is 
\Eq{
N=\frac{1}{2\lambda t +c},\quad
q_{ij}=|2\lambda t + c| q_{ij}^0(z).
}
In terms of $\tau$ defined by
\Eq{
|2\lambda t + c|=c_1 e^{2\lambda\tau},
}
this metric can be put into the form
\Eq{
ds^2=c_1^2 e^{2\lambda \tau}\insbra{\pm d\tau^2+q_{ij}^0(z)dz^idz^j},
}
where $\pm$ is the sign of $2\lambda t+c$. For this metric, the 
condition $R_{\mu\nu}({\rm X}_4)=0$ is equivalent to
\Eq{
R_{ij}(q^0)=\pm 2\lambda^2 q^0_{ij}.
}
We can show by direct calculations that $ds^2({\rm X}_4)$ is flat under 
this condition. 

Next, we consider the case in which $Dh_0\not=0$ and $(Dh_0)^2=0$. 
In this case, from $D_\mu (Dh_0)^2=2D^\nu h_0 D_{\mu}D_{\nu}h_0=0$ it 
follows that $\lambda$ should vanish. Further, $u=h_0$ is a null 
coordinate and $Du$ becomes a null Killing. Therefore, the metric 
can be written
\Eq{
ds^2=2du\inpare{d\rho+a du + b_i dz^i}+q_{ij}dz^i dz^j,}
where $b_i$ and $q_{ij}$ are functions independent of $\rho$.  
If we define the null vectors $e_\pm$ as
\Eq{
e_+=\partial_u -a \partial_\rho,\quad
e_-=\partial_\rho,
}
the Ricci curvature is expressed as
\Eqrsub{
&& R_{-\mu}=0,\\
&& R_{++}=-\triangle a + D^i(\partial_u b_i)-\partial_u K
   -K_{ij}K^{ij}+2B^2,\\
&& R_{+i}=-\partial_i K + D_j K^j_i + \epsilon_{ij}D^j B,\\
&& R_{ij}=R_{ij}(q),
}
where
\Eq{
K_{ij}=\frac{1}{2}\partial_u q_{ij},\quad
B=\frac{1}{2}\epsilon^{ij}\partial_{[i} b_{j]}.
}

First, from the condition $R_{ij}(q)=0$, we can set 
$q_{ij}=\delta_{ij}$ by an appropriate coordinate transformation. In 
this coordinate system, $K_{ij}=0$. Hence from  $R_{+i}=0$ we have 
$B=B(u)$. This implies that $b_i$ can be written
\Eq{
b_i=-B\epsilon_{ij}z^j + \partial_i C(u,z)\,.
}
Inserting this to $R_{++}=0$, we obtain
\Eq{
\triangle_2 (-a+ \partial_u C)+2B^2=0,
}
whose general solution is
\Eq{
a=\partial_u C +\frac{B^2}{2}z^i z_i +\frac{f}{2},
}
where $f$ is an arbitrary solution to $\triangle_2 f=0$. Thus, after 
the coordinate transformation $\rho+C\maps \rho$, the metric can be 
expressed as
\Eq{
ds^2=2du\,d\rho+ fdu^2
 +(dz^1 - Bz^2 du)^2 + (dz^2+B z^1 du)^2.
}
The only non-vanishing components of the curvature tensor for this 
metric are 
\Eq{
R_{u i u j}=-\frac{1}{2}\partial_i\partial_j f.
}
Therefore, if $f$ is linear with respect to $z^i$, the spacetime is 
flat. For example, in terms of the coordinates defined by
\Eqrsub{
&& t=\frac{1}{2}u-\rho,\quad
   z=-\frac{1}{2}u -\rho,\\
&& x=z^1 \cos\beta + z^2 \sin\beta,\quad
   y=-z^1 \sin\beta +z^2\cos\beta,
}
where $\beta=\int B du$, the metric can be written
\Eq{
ds^2=-dt^2+dx^2+dy^2+dz^2+ f(x,y,t-z)(dt-dz)^2,
}
where $f(x,y,t-z)$ is a harmonic function with respect to $x$ and 
$y$.

%T2>Deformed conifold solutions
\section{General solution for the deformed conifold}
\label{Appendix:deformed}
In this appendix, we give the general solution for the deformed 
conifold case discussed in \S\ref{subsec:deformed}.

First, the general solution for $F$ , $k_1$  and $k_2$ are given by 
\begin{subequations}
\begin{eqnarray}
F&=&\frac{1}{2} \left(1-\frac{\tau}{\sinh\tau}\right)
     +\frac{C_1}{\sinh\tau}
     +C_2 \frac{2\tau-\sinh(2\tau)}{2\sinh(\tau)}\,,
       \\
k_1&=&C_0+\alpha \frac{1-e^{2\tau}+\tau (1+e^{2\tau})}{2(1+e^{\tau})^2} 
    + \alpha C_1 \frac{2e^{\tau}}{(1+e^{\tau})^2}
        \nn\\
   & &+\alpha C_2 \left\{\frac{e^{3\tau}+e^{2\tau}-5e^{\tau}-1}
       {2 e^{\tau} (1+e^{\tau})}
      -2\tau \frac{e^{2\tau}+e^{\tau}+1}{(1+e^{\tau})^2}\right\}\,,
       \\
k_2&=&C_0+\alpha \frac{1-e^{2\tau}+\tau (1+e^{2\tau})}
      {2 (1-e^{\tau})^2}
      -\alpha C_1 \frac{2e^{\tau}}{(1-e^{\tau})^2}
       \nn\\
    & &-\alpha C_2 \left\{\frac{e^{3\tau}-e^{2\tau}-5e^{\tau}+1}
       {2 e^{\tau} (e^{\tau}-1)}
      +2\tau \frac{e^{2\tau}-e^{\tau}+1}{(e^{\tau}-1)^2}\right\}\,.
\end{eqnarray}
\end{subequations}
For the deformed conifold, \eqref{Sol:general:h1} reads
\Eq{
\frac{1}{\sinh^2\tau} \pd_{\tau} 
     \left(\frac{6K^2 \sinh^2\tau}{\sigma^3} \pd_{\tau} h\right)
    =-g_s^{-1} (H_3\cdot H_3)_Y\,,
}
where $(H_3\cdot H_3)_Y$ in the right-hand side is given by
\Eq{
H_3\cdot H_3=\frac{24 g_s^2 \alpha^2}{\sigma^9} 
      \left[\frac{(1-F)^2}{\cosh^4(\tau/2)}
     +\frac{F^2}{\sinh^4(\tau/2)}
     +\frac{2(k_1-k_2)^2}{\alpha^2 \sinh^2\tau}\right]\,.
}
Hence, the general solution for the warp factor $h$ can be obtained 
by integrating
\begin{eqnarray}
\pd_{\tau} h&=&-\frac{16\cdot 2^{2/3} g_s \alpha^2}
       {\sigma^6 (\sinh(2\tau)-2\tau)^{2/3} \sinh^3\tau}\,
       \nn\\
    & &\times \Big[-\frac{1}{4} (\sinh(2\tau)-2\tau) 
        (\tau \cosh\tau-\sinh\tau)
       +2C_1(\tau\cosh\tau-\sinh\tau)
       \nn\\
   & &+C_2 \Big\{2\tau^2 \cosh\tau
       -\frac{\tau}{2} (\sinh(3\tau)+5\sinh\tau)
       +e^{-\tau} \sinh\tau \Big\}
       \nn\\
   & &-4C_1 C_2 (\tau \cosh\tau-\sinh\tau)-8C_1^2 \cosh\tau
       \nn\\
   & &+C_2^2 \Big\{-4\tau^2 e^{2\tau} \cosh\tau 
      +4\tau e^{2\tau} \sinh\tau
       \nn\\
   & &+\frac{1}{8} \left(e^{7\tau}-3e^{5\tau}
     -6e^{3\tau}+18e^{\tau}-11e^{-\tau}+e^{-3\tau}\right)\Big\}
       \nn\\
   & & +C_3 \sinh^3\tau \Big]\,.
\end{eqnarray}
%

%T1>References
%======================================%
%<<<<<<<<<<<<< REFERENCE >>>>>>>>>>>>>>%
%======================================%

\end{document}